# A Family of Pb-based Superconductors with Variable Cubic to Hexagonal Packing

Tai Kong[1], Karolina Górnicka[2], Sylwia Gołąb[3], Bartlomiej Wiendlocha[3]
Tomasz Klimczuk[2] and Robert. J. Cava[1]

[1] *Department of Chemistry, Princeton University, Princeton, NJ 08544, USA*
[2] *Faculty of Applied Physics and Mathematics, Gdansk University of Technology, Narutowicza 11/12, 80-233 Gdansk, Poland*
[3] *Faculty of Physics and Applied Computer Science,*
*AGH University of Science and Technology, Aleja Mickiewicza 30, 30-059 Krakow, Poland*

**Abstract**

We describe three previously unreported superconductors, $BaPb_3$, $Ba_{0.89}Sr_{0.11}Pb_3$ and $Ba_{0.5}Sr_{0.5}Pb_3$. These three materials, together with $SrPb_3$, form a distinctive isoelectronic family of intermetallic superconductors based on the stacking of Pb planes, with crystal structures that display a hexagonal to cubic perovskite-like progression, as rarely seen in metals. The superconducting transition temperatures ($T_c$) are similar for all - 2.2 K for $BaPb_3$, 2.7 K for $Ba_{0.89}Sr_{0.11}Pb_3$ and 2.6 K for $Ba_{0.5}Sr_{0.5}Pb_3$, and the previously reported $T_c$ of $SrPb_3$, ~ 2 K, is confirmed. The materials are moderate coupling superconductors, and calculations show that the electronic densities of states at the Fermi energy are primarily contributed by Pb. The observations suggest that the Pb-stacking variation has only a minor effect on the superconductivity.

**Introduction**

The crystal and electronic band structures of a material are two of the most important factors that influence its electronic properties. In terms of superconductivity, the large majority of studies focus on tuning a system within the same crystal structure through changing the electron count. This allows one to vary the density of states at the Fermi level and the topology of the Fermi surface and has been documented many times. For simple binary metallic alloy superconductors, for example, Matthias rules, based on a rigid band picture of *d*-orbital filling, relate electron count and superconducting transition temperatures[1,2], while for more complex superconductors such as the cuprates or the pnictides, the Fermi surface topology and its changes with electron count are at the center of most discussions about the mechanisms for superconductivity[3].

Crystal structure also plays a key role in the properties of superconductors. High crystal symmetries appear to favor conventional superconductivity[4] while for high temperature superconductors, structural motifs such as layers of Fe-As tetrahedra or $CuO_4$ squares are important[5,6,7]. While it is relatively easy to tune electron count in a given material, it is much more difficult to find a system where the crystal structure can be systematically varied while maintaining the same electron count and electronically-active elemental constituents. The most common method for doing so is via the application of pressure, which compresses crystal structures, changing the interactions of atoms' atomic orbitals and therefore their hybridization, and, as a consequence, the electronic bandwidth, the Fermi surface, and the electron-phonon coupling. Pressure can also sometimes drive a structural phase transition - in the case of $CaFe_2As_2$, for example, a collapse of the tetragonal *c* axis under pressure has been associated with the destruction of superconductivity[8,9]. In superconductors like the pnictides and the cuprates, the spacer layers between isoelectronic electronically active layers can be varied and the impact on the superconductivity tested[5,6,7].

Examples of structural families that maintain atomic ratios, electron count, and electronically active elements while systematically varying the crystal structure are rare. The most well-known structurally evolving family of this type may be the $ABO_3$ cubic to hexagonal oxide perovskites. Depending on the ionic size ratio of atoms involved, different types of packing of the transition metal-oxygen ($BO_6$) octahedra, and the associated oxygen layers, are found[10]. In the $SrRuO_3$ to $BaRuO_3$ series, for example, the crystal structure gradually evolves from cubic packing on the Sr-rich side to more hexagonal packing on the Ba-rich side[11] – Sr being a smaller A site ion than Ba. Between the purely cubic structure and the purely hexagonal structure, various ordered combinations of cubic and hexagonal packing are found when Ba and Sr are randomly mixed and the average size of the atom on the A site is varied. When, in addition to the Ru, a second metal is also permitted, the possibilities for different ratios of cubic to hexagonal stacking appear to be endless. Unfortunately, superconductivity is not known to exist in this family.

In contrast, in intermetallic compounds such Perovskite-like structural evolution is extremely rare. It has been shown to exist, however, in the $SrPb_3$–$BaPb_3$, $SrSn_3$-$BaSn_3$ and $BaSn_3$-$BaBi_3$ families[12,13,14] (they are intermetallic perovskites with the "B" site empty) where an evolution from cubic to hexagonal packing has been observed as a function of the size of the atoms (i.e. Sr and Ba) on the "A" site. Because some such materials are reported superconductors[15,16,17], the one relevant to the current work being $SrPb_3$ reported to superconduct below 1.85 K[18,19], these compounds potentially offer a unique opportunity for the systematic study of superconductivity within a structural family without changing the electron count or the chemical character of the elements involved (Sr and Ba are in the same column of the periodic table). Fig. 1 shows a schematic drawing of the structural evolution of the $SrPb_3$ – $BaPb_3$ intermetallic family compared to an analogous family of oxide materials. In $SrPb_3$ and $SrRuO_3$, the $\square Pb_6$ or $RuO_6$ octahedra are connected by sharing corners, resulting in a cubic (or tetragonal or orthorhombic, depending on octahedron tilts) structure. With increasing fraction of the larger Ba ions, face sharing of the octahedra is introduced, and, eventually, this becomes the primary structural motif seen in $BaPb_3$ or $BaRuO_3$. The Pb or O arrays change from cubic packing to hexagonal packing in the two pure end members. Between the end member compounds, the cubic packing (corner sharing of octahedra) and hexagonal packing (face sharing of octahedra) are both present, depending on the average size of the atoms, yielding distinct compounds that exhibit different crystal structures. The structural ordering in this kind of family is denoted by the sequence of corner sharing (c) and face sharing (f) octahedra in the crystal structure. For example in Fig. 1(b), the packing order can be denoted as ccfccf…, or a 33% hexagonal packing. In a similar manner, the packing sequence in $Ba_{0.5}Sr_{0.5}Pb_3$ can be denoted as ccfccf (33% hexagonal packing) and $Ba_{0.89}Sr_{0.11}Pb_3$ can be denoted as ccffccff (50% hexagonal packing). Figure 1 also presents examples of oxides outside of the $SrRuO_3$-$BaRuO_3$ family where such motifs are known.

Here we focus on the $SrPb_3$-$BaPb_3$ family of materials. We report the discovery of superconductivity in the $BaPb_3$, $Ba_{0.89}Sr_{0.11}Pb_3$ and $Ba_{0.5}Sr_{0.5}Pb_3$, members of this structural family, and compare their properties to that of $SrPb_3$. The ratio of cubic to hexagonal packing of the Pb atoms varies in a systematic fashion in the series. A variation in $T_c$ is observed, with one of the intermediary materials showing a higher $T_c$ than the others. Overall, however, the superconductivity appears to be relatively insensitive to the variation from cubic to hexagonal Pb packing in this family. We speculate that to first order, these materials appear to be superconducting Pb with different lattice geometries.

**Methods**

BaPb$_3$ and Ba$_{0.89}$Sr$_{0.11}$Pb$_3$ single crystals were grown out of a molten Pb solution. Starting elements were packed in an alumina crucible in an atomic ratio of A:Pb = 15:85. (A = alkaline earth element Ba and/or Sr.) The packed materials were then sealed in an evacuated silica tube, heated to 650 $^o$C, and slowly cooled to 360 $^o$C, at which temperature the molten solution was decanted in a centrifuge. SrPb$_3$ and Ba$_{0.5}$Sr$_{0.5}$Pb$_3$ were obtained by direct melting of the stoichiometric elemental mixtures in an alumina crucible inside an evacuated silica tube. These materials were heated to 800 $^o$C and then quenched to room temperature. All of the resulting materials are silvery-colored and air-sensitive.

Powder x-ray diffraction measurements were performed using a Bruker D8 Advance Eco, Cu K$_\alpha$ radiation (λ=1.5406 Å), equipped with a LynxEye-XE detector. A kapton film was used to cover the powder samples to reduce air exposure during data acquisition. Magnetization measurements were performed using a Quantum Design Physical Property Measurement System (PPMS) Dynacool equipped with a VSM function. Samples were loaded in a plastic capsule inside of the glove box and quickly transferred to the machine for measurements. Both zero field cooled (ZFC) and field cooled (FC) data were collected. Heat capacity measurements were carried out in a PPMS Evercool II system by using a 2τ relaxation method. Due to the presence of filamentary superconductivity from remnant Pb, electrical resistivity data were not considered reliable and are not reported.

Band structure calculations were performed using the full-potential linearized augmented plane-wave method with local orbitals (LAPW+LO) and the Perdew–Burke–Ernzerhof generalized gradient approximation (PBE-GGA) exchange-correlation potential, as implemented in the WIEN2k package[21]. Scalar-relativistic (without the spin-orbit coupling, SOC) and relativistic (including SOC, with relativistic local orbitals added for all atoms) calculation results are presented, to highlight the importance of the spin-orbit interaction in the electronic structures. In all cases, calculations were done using the experimentally determined crystal structures[12]. For the mixed Sr-Ba-based compounds, the supercell technique was used. For the Sr$_{0.5}$Ba$_{0.5}$Pb$_3$ case, the supercell was constructed by occupation of half of the positions with Sr and half with Ba in one unit cell containing six formula units (Sr$_3$Ba$_3$Pb$_{18}$, space group changed from *P6$_3$/mmc* (no. 194) to *P-3m1* (no. 164) to accommodate the supercell, with the lattice parameters unchanged: a=7.271 Å, and c/a=4.703. Two distributions of Sr and Ba atoms in the unit cell were tested and the one with lower total energy was chosen for the final calculations. For Sr$_{0.11}$Ba$_{0.89}$Pb$_3$, the supercell contained 12 formula units (space group changed from *R-3m* (no. 166) to *P3m1* (no. 156) to accommodate the supercell, with lattice parameters a=7.156 Å, and c/a=2.400 with one site occupied by Sr, and 11 by Ba, resulting in a final $SrBa_{11}Pb_{36}$ structure, i.e. $Sr_{0.083}Ba_{0.917}Pb_3$ stoichiometry, slightly different from the experimentally studied one. To verify the supercell approach and especially to check whether the small difference of chemical composition between the $Sr_{0.083}Ba_{0.917}Pb_3$ supercell, employed in the calculation, and the actual $Sr_{0.11}Ba_{0.89}Pb_3$ composition was significant, we used the Korringa-Kohn-Rostoker method with the coherent potential approximation (KKR-CPA, in the Munich SPR-KKR package[22,23]). KKR-CPA calculations, which explicitly take into account the chemical disorder on the Ba/Sr sublattice and allowed to compute the exact $Sr_{0.11}Ba_{0.89}Pb_3$ composition, showed no important differences in the DOS values near the Fermi energy, when comparing to the supercell results.

**Results**

The powder x-ray diffraction data for our samples of the members of the $SrPb_3$-$BaPb_3$ series of compounds are shown in Fig. 2. LeBail fits (red lines) were performed to fit the experimental data shown in black crosses. All peaks can be indexed by either members of the target family[12] or elemental Pb, confirming a good agreement between the compounds synthesized in this study with those previously reported.

The superconducting behavior was first studied via magnetization measurements. The temperature-dependent magnetization data for the $SrPb_3$- $BaPb_3$ family of materials are shown in Fig. 3, plotted as (negative) volume magnetic susceptibility versus temperature. Both ZFC and FC data were collected at 10 Oe. (The obtained volume fractions exceed -1 because we did not apply a demagnetization correction.) The bulk nature of these superconducting transitions is confirmed by our heat capacity measurements, as shown below. For all four compounds, we observed minor amounts of diamagnetism onsetting at 7.2 K, a result of the presence of elemental Pb either from a small amount of decomposition of the sample in air, or from remnant flux. The superconducting transition temperatures in the magnetization measurements were determined conservatively, from the intersection point of the linearly extended normal state magnetization and the line tangent to the steepest drop in susceptibility at the superconducting diamagnetic transition[24]. These numbers are indicated by dashed lines in Fig. 3. The superconducting $T_c$ for $SrPb_3$ is close to 2 K, slightly higher than the reported value of 1.85 K[18,19]. The $T_c$s for the previously unreported superconductors $Ba_{0.5}Sr_{0.5}Pb_3$, $Ba_{0.89}Sr_{0.11}Pb_3$ and $BaPb_3$ are 2.5 K, 2.7 K and 2.1 K respectively. The difference in ZFC and FC magnetizations, albeit clear, is smaller in $BaPb_3$ and $Ba_{0.89}Sr_{0.11}Pb_3$ than in $SrPb_3$ and $Ba_{0.5}Sr_{0.5}Pb_3$, an indication of their different flux-pinning characteristics: the single crystalline samples of $BaPb_3$ and $Ba_{0.89}Sr_{0.11}Pb_3$ are not as efficient at pinning magnetic flux as are the polycrystalline samples of $SrPb_3$ and $Ba_{0.5}Sr_{0.5}Pb_3$.

The nature of the superconductors was further characterized by temperature dependent specific heat measurements (Fig. 4). The $T_c$s are determined from the commonly employed equal entropy criterion (shown by dashed lines in Fig. 4) so that the total entropy is conserved when comparing the slightly rounded experimental data with an ideally sharp jump at $T_c$. The superconducting $T_c$s thus obtained are 2.0 K, 2.6 K, 2.7 K and 2.2 K for $SrPb_3$, $Ba_{0.5}Sr_{0.5}Pb_3$, $Ba_{0.89}Sr_{0.11}Pb_3$ and $BaPb_3$. These values are consistent with those obtained in the temperature dependent magnetization measurements.

We can also estimate the electronic contribution to the specific heat, $\gamma$, and the Debye temperature, $\Theta$, from the fits to the low temperature data in $C_p/T$ versus $T^2$ plots. The obtained values, together with the $T_c$s, are listed in Table I. Generally, the $\gamma$ values are around 5 mJ/mol-$K^2$, or 1 mJ/mol-atom-$K^2$. This value is close to that of a normal metal without a significant mass enhancement[25]. At the superconducting transition temperature, however, we observe relatively large jumps in specific heat. The specific heat jump ratio, $\alpha = \Delta C/\gamma T_c$, ranges from 1.5 for $SrPb_3$ to 2.0 for $Ba_{0.89}Sr_{0.11}Pb_3$. These values are larger than what is expected for weak coupling superconductors, 1.43, especially for $Ba_{0.89}Sr_{0.11}Pb_3$. (Elemental Pb also displays a relatively large $\alpha$, around 2.7 [26,27].) We note that although the superconducting Pb impurity is seen in the magnetization data, the signature of Pb superconductivity is very subtle in the specific heat, indicating that it is present in only small amounts in the samples, and thus does not contribute significantly to the observed large specific heat jumps shown in Fig. 4. Our experimentally determined $\gamma$ values for the new superconductors are therefore only weakly overestimated due to potential Pb contamination. The Debye temperatures are around 130 K for all compounds. A summary of the variation of the superconducting transition temperatures $T_c$, Sommerfeld coefficients $\gamma$ and specific heat jump ratios $\alpha$ is presented in Figure 5.

With the application of external magnetic field, the superconducting transitions are broadened and the $T_c$s are suppressed. The critical fields needed to suppress the superconductivity are relatively low for the end members: only 40 Oe is required to suppress the $T_c$ below 1.8 K (the lower temperature limit of our magnetization measurements) for $SrPb_3$ and 100 Oe for $BaPb_3$. For the middle two members in the series, however, it takes a significantly larger magnetic field to suppress the superconductivity. The critical magnetic field at 1.8 K is around 1 kOe for both $Ba_{0.89}Sr_{0.11}Pb_3$ and $Ba_{0.5}Sr_{0.5}Pb_3$. Given very similar $T_c$ and $\gamma$ values, the difference in critical field value may originate from a difference in their anisotropic superconducting gap structure[28].

Figure 6 shows the computed densities of states (DOS) for all the materials in the family (a-d shows the scalar-relativistic results, e-h and i-l shows the relativistic ones), and Figure 7 shows the electronic band structures and Fermi surfaces (FS) for the end-member compounds, SrPb$_3$ and BaPb$_3$. The values of DOS at the Fermi level are shown in Table 1.

Our scalar-relativistic results are consistent with those previously reported in Ref. 12. In all the compounds there are 14 valence electrons per formula unit (f.u.): each Pb atom contributes two 6*s* and two 6*p* electrons, while the Sr and Ba atoms contribute two 5*s* or 6*s* electrons, respectively. The *6s* electronic orbitals of Pb form a separate DOS block in the energy range from -12 to -6 eV, as they are more strongly bound to the core. The main valence band is built up from the hybridized 6*p* orbitals of Pb and the 5*s* (or 6*s*) orbitals Sr (or Ba) in the energy range from -4.5 eV to the Fermi level. (The differences in the number of electronic bands in Figs. 6(a) and 6(b) originates in the different number of formula units in the primitive unit cell of SrPb$_3$ and BaPb$_3$ - in the tetragonal cell of SrPb$_3$ there is one f.u. per cell, while in the rhombohedral cell of BaPb$_3$ there are 3 f.u., thus there are 3 times more bands.) The DOS of BaPb$_3$ and the Sr/Pb intermediate alloys are quite similar to each other. The DOS of SrPb$_3$ is different, however; the lower part of the DOS, which comes from *s*-states, is more similar to the DOS of *fcc* Pb. In all cases, the minimum in the DOS near the Fermi level is created and is deepened as the Sr atom fraction increases.

In SrPb$_3$, (Fig. 7a) two bands with Pb-*p* character, parabolic-like in the vicinity of the A-point in the Brillion Zone, cross the Fermi level, resulting in two pieces of Fermi surface (Fig. 7b-c). The shape of the Γ-centered one with flat planes is quite similar to the one of the Fermi surface pieces in elemental Pb (Fig. 7d) and is not affected by spin-orbit coupling. The second part of the Fermi surface also resembles that for Pb, consisting of tubes centered along M-X, X-R, and Z-R. This part of the Fermi surface is more sensitive to SOC – in the calculations with SOC included it changes into series of pockets associated with a pseudogap that opens in the M-X-R-Z directions.

In BaPb$_3$, in contrast, three bands cross the Fermi level (Fig. 7e), resulting in three sheets of Fermi surface (Fig. 7f-g). The first piece is electron-like and consists of two cones centered at the T point of the Brillouin Zone. It is associated with a parabolic-like band in the B-T-C plane, which becomes linear along the trigonal axis (the Γ-T direction). The second piece of the Fermi surface consist mainly of two pockets – one centered at the Γ point and another centered at the T point, associated with a band similar to the first one in the vicinity of the T point. These first and second bands are degenerate at the T point. A third, hole-like piece consists of two flat pockets centered along the trigonal axis (but not at the Γ point) and six half-rings connected to each other. Only the first, cone-like sheet of FS is not sensitive to the spin-orbit coupling, the two remaining ones are strongly affected by SOC.

It is worth noting here the similarities between the band structures of $SrPb_3$ and $BaPb_3$ around the A (and T) points, which are not accidental. This is because the A point is located on the diagonal of the tetragonal BZ of $SrPb_3$, which becomes the trigonal axis in the rhombohedral unit cell of $BaPb_3$, where the T-point is found. Both of these directions are related to the real-space Pb-layers stacking direction (see. below).

The densities of states calculations show that when the SOC is included, a decrease in the DOS at the Fermi level occurs (see, Table 1). Although the DOS functions change overall when the Sr to Ba ratio is varied, the (SOC-included) DOS values at the Fermi level remain fairly constant in the series. Assuming that the electron-phonon (*e-p*) interaction renormalizes the electronic specific heat, the magnitude of the *e-p* interaction can be estimated by comparing the band-structure-calculated value of the Sommerfeld coefficient $\gamma_{theo}$ and the experimentally determined one via: $\gamma_{expt} = \gamma_{theo}(1+\lambda_{ep})$. The calculated values, presented in Table 1, are close to the values determined using the McMillan formula[29] for $T_c$:

$$\lambda_{ep} = \frac{1.04 + \mu^* ln(\frac{\Theta}{1.45T_c})}{(1 - 0.62\mu^*)ln(\frac{\Theta}{1.45T_c}) - 1.04}$$

where the Coulomb pseudopotential, $\mu^*$ is fixed to the commonly used value, 0.13. The calculated $\lambda_{ep}$ values can be considered as reflecting an intermediate coupling strength[29].

**Discussion**

As previously described, Figure 5 shows the superconducting transition temperature, $T_c$, electronic specific heat, $\gamma$, and specific heat jump ratio, $\alpha$, as a function of hexagonal packing ratio. For both the $T_c$ and $\alpha$ values, there is a broad maximum in the middle of the series, a reflection of larger electron-phonon coupling. For comparison, the related material $SrBi_3$ has been identified as displaying enhanced electron-phonon coupling[30] based on its $\alpha$ value, while, $\alpha$ is around 1.02 in $BaBi_3$, much lower than the weak coupling BCS estimated value[15].

Information regarding the chemical bonding and the role of the Sr/Ba atoms in determining the crystal structures of these materials can be obtained by an analysis of the charge density. In Fig. 8, various projections of the valence charge density (i.e. corresponding only to the bands between -4.5 eV and the Fermi level) are shown for $SrPb_3$ (panel a), $Ba_{0.5}Sr_{0.5}Pb_3$ (panel b) and $BaPb_3$ (panel c). In all cases the layered stacking of Kagome or distorted Kagome sheets, oriented perpendicular to the c-axis of the hexagonal lattices, or to the {111} axis in the case of $SrPb_3$, can be identified. Pb Kagome nets are the main building blocks of these layers, and are easily found in Fig. 8 - in most layers, these are centered by the alkaline earth element. This type of view offers an alternative way of looking at these structures as variations of the face centered cubic structure of metallic lead, where, the hexagonal layers are stacked along the {111} axis of the unit cell.

In this picture, in $SrPb_3$ and $BaPb_3$ one quarter of the atoms are Sr or Ba and the remainder mimic Pb in metallic *fcc* lead. Sr is larger than Pb, and thus $SrPb_3$ has a slightly larger molar volume per atom than *fcc* Pb does. The tetragonal unit cell of $SrPb_3$ is then only a slightly distorted cubic structure of Pb, since the *a, b* (4.96 Å) and *c* parameters (5.02 Å) are similar and close to the lattice parameter of *fcc* Pb (4.95 Å). Further, the distance between the Pb atoms in the kagome {111} planes is very similar to that in metallic lead (3.5 Å). Although there are two crystallographically inequivalent positions in $SrPb_3$, all the Pb atoms contribute to form a single Pb sublattice. As Sr is an electropositive element, charge is transferred from Sr towards Pb, in an amount equal to about 0.5 e per Sr atom, showing that there is partially ionic bonding between Sr and the Pb in the structure. In contrast, the sublattice of Pb forms a metallic bonding array, with only a small charge transfer, of 0.01 $e/a_B^3$, between them.

The situation becomes more complicated when Ba is introduced. For $BaPb_3$ the distortion from the cubic *fcc* lattice is stronger due to the larger size of Ba, but the sequence of the hexagonal-like layers, now stacked along the trigonal axis, is maintained. In the hexagonal unit cell of $BaPb_3$, two types of layers can be distinguished, in an ABB'A'B''-like sequence. The top (A) layer, with the charge density plotted in Fig. 8c, consists of Pb Kagomes, centered on the Ba atom. These Kagomes are expanded compared to those in $SrPb_3$ and metallic Pb, with Pb-Pb (and Pb-Ba) distances equal to 3.64 Å. The next, B, layer is a distorted one, with Pb and Ba atoms slightly moved out of plane. This layer has a different charge density distribution, as seen in figure 8c, and in this plane there is a dichotomy in the Pb bonding - the Pb equilateral triangles are either smaller (3.21 Å), or larger (4.07 Å) than the average. In contrast, almost no change in the Pb-Ba distance is observed - here it is equal to 3.65 Å. The next B' layer is a translated mirror reflection of B, and similarly the next layers are a translated layer A and a mirror reflection of B - two such sequences form the rhombohedral unit cell. For $BaPb_3$ we have a charge transfer from Ba to Pb that is similar yet smaller than that between Sr and Pb in $SrPb_3$.

In the intermediate members $Ba_{0.11}Sr_{0.89}Pb_3$ and in $Ba_{0.5}Sr_{0.5}Pb_3$ a layer scheme similar to that of $BaPb_3$ is found, although number of layers is different in these cases. Layers with shorter Ba-Pb and Sr-Pb distances are present in these materials, 3.64 Å in $Ba_{0.11}Sr_{0.89}Pb_3$ and 3.58 Å in $Ba_{0.5}Sr_{0.5}Pb_3$ case. One can see hexagonal-like charge density in these layers and also in the layers with shorter Pb-Pb distances (3.16 Å in both cases), where charge density is rather triangle-like.

Finally, we note that a similar series, although previously unrecognized as such, also exists for $BaSn_3$, $SrSn_3$ and $CaSn_3$. The superconducting transition temperatures are reported as 2.4 K for $BaSn_3$[16], 5.4 K for $SrSn_3$[17], and 4.2 K for $CaSn_3$[31]. Thus this series of compounds also shows a maximum in $T_c$ in the middle of the structural sequence. The present results suggest that other members in the Ca/Ba/SrSn$_3$ family may also be superconducting. In a hypothetical $BaSn_3$-$SrBi_3$ or $BaPb_3$ – $SrBi_3$ series, where both electron count and crystal structure would be changed[13], more complex behavior may be expected.

**Conclusions**

We report the discovery of superconductivity in $BaPb_3$ (2.2 K), $Ba_{0.89}Sr_{0.11}Pb_3$ (2.7 K) and $Ba_{0.5}Sr_{0.5}Pb_3$ (2.6 K). These compounds have perovskite-like structures based on $Pb_6$ octahedra. Specific heat measurements show large jumps at the superconducting transition temperatures, indicating that these are moderate coupling strength superconductors. The calculated densities of states, which are sensitive to the presence of spin orbit coupling, are close to the experimentally obtained values. Despite having distinct crystal structures, we speculate that the superconductors in the $BaPb_3$-$SrPb_3$ family can be considered as superconducting Pb with different layer stacking and small amount of charge transfer from the Ba/Sr atoms to the Pb sublattice. Although the $T_c$s are lower than Pb, the materials appear to be moderately strongly coupled BCS superconductors. Further study of this family of intermetallic perovskites and the related Sn-based family would be of future interest.

**Acknowledgements**

This work was supported by the Gordon and Betty Moore EPiQS initiative, grant number GBMF-4412. Research performed at the AGH-UST was supported by the National Science Center (Poland), project No. 2017/26/E/ST3/00119. SG was partly supported by the EU Project POWR.03.02.00-00-I004/16.

TABLE I. Characteristic values for Ba/SrPb$_3$ superconductors: superconducting transition temperature, $T_c$; both experimental and theoretical electronic specific heat, $\gamma$; Debye temperature, $\Theta$; specific heat jump ratio at $T_c$, $\alpha$; both experimental (based on McMillan formula) and theoretical (based on renormalization of $\gamma$) electron-phonon coupling constant, $\lambda_{ep}$ and density of states (DOS) at the Fermi energy, for the scalar-relativistic (SR) and relativistic (SOC) calculations.

|  | SrPb$_3$ | Ba$_{0.5}$Sr$_{0.5}$Pb$_3$ | Ba$_{0.89}$Sr$_{0.11}$Pb$_3$ | BaPb$_3$ |
|---|---|---|---|---|
| $T_c$ (K) | 2.0 | 2.6 | 2.7 | 2.2 |
| $\Theta$ (K) | 135(1) | 127(1) | 132(1) | 133(1) |
| $\alpha = \Delta C / \gamma T_c$ | 1.5 | 1.6 | 2.0 | 1.8 |
| $\gamma$ (mJ/mol-K$^2$)(exp) | 5.3(2) | 5.7(4) | 5.7(4) | 5.3(2) |
| $\gamma$ (mJ/mol-K$^2$)(theo) | 3.76 | 3.98 | 4.10 | 3.97 |
| $\lambda_{ep}$ (exp) | 0.62 | 0.69 | 0.69 | 0.64 |
| $\lambda_{ep}$ (theo) | 0.66 | 0.43 | 0.39 | 0.33 |
| DOS (/eV-f.u.)(exp) | 1.39 | 1.44 | 1.44 | 1.38 |
| DOS (/eV-f.u.)(theo-SR) | 1.59 | 1.77 | 2.73 | 1.94 |
| DOS (/eV-f.u.)(theo-SOC) | 1.36 | 1.69 | 1.74 | 1.69 |

**Figure captions**

FIG. 1. (Color online) A comparison between crystal structures of $SrRuO_3$ - $BaRuO_3$ and $SrPb_3$ - $BaPb_3$ with a cascade change of packing types. The upper row shows the structure of (a) $SrRuO_3$, (c) 6-layer (6L) Ba/SrRuO$_3$, and (d) $BaRuO_3$[11]. The lower row shows the structures of (e) $SrPb_3$, (f) $Ba_{0.5}Sr_{0.5}Pb_3$, (g) $Ba_{0.89}Sr_{0.11}Pb_3$ and (h) $BaPb_3$[12]. Only the $RuO_6$ and $\square Pb_6$ octahedra are shown for clarity. The structure of $Ba_{0.89}Sr_{0.11}Pb_3$ has no strict analog in the $SrRuO_3$-$BaRuO_3$ family, but has been observed for (c) $Ba_4LaRu_3O_{12}$[20].

FIG. 2. (Color online) Powder x-ray diffraction data for (a) $SrPb_3$, (b) $Ba_{0.5}Sr_{0.5}Pb_3$, (c) $Ba_{0.89}Sr_{0.11}Pb_3$ and (d) $BaPb_3$. Black crosses are measured data and red lines are LeBail fitted curves. The magenta ticks at the bottom are Bragg reflection positions for each structure.

FIG. 3. (Color online) ZFC and FC magnetization data measured under 10 Oe for (a) $SrPb_3$, (b) $Ba_{0.5}Sr_{0.5}Pb_3$, (c) $Ba_{0.89}Sr_{0.11}Pb_3$ and (d) $BaPb_3$. Insets show expanded views on the low-temperature superconducting transition of each compound. The superconducting transition temperatures were inferred from the cross point of the dashed lines (see texts).

FIG. 4. (Color online) Temperature and field dependent specific heat data for (a) $SrPb_3$, (b) $Ba_{0.5}Sr_{0.5}Pb_3$, (c) $Ba_{0.89}Sr_{0.11}Pb_3$ and (d) $BaPb_3$. Insets show linear fits to low-temperature $C_p/T$ versus $T^2$ plots to extract the electronic specific heat γ and Debye temperature Θ.

FIG. 5. (Color online) A summary of superconducting transition temperature, $T_c$, electronic specific heat, γ and specific heat jump ratio, $\alpha = \Delta C/\gamma T_c$ as a function of hexagonal packing ratio (see text).

FIG. 6. (Color online) Density of states of Ba/SrPb$_3$ compounds calculated without (a-d) and with (e-h) spin-orbit coupling (SOC), and using supercell models for the two middle cases. Panels (i-l) show the results with SOC in the vicinity of Fermi level.

FIG. 7. Band structure of $SrPb_3$ (a) and $BaPb_3$ (e) calculated without SOC (red line, "scalar") and with SOC (black line, "relativistic", "rel."). The Fermi surfaces of $SrPb_3$ (b, c) and $BaPb_3$ (f, g) without and with SOC. The Fermi surface of $SrPb_3$ is quite similar to that of *fcc* Pb (d).

FIG. 8. Charge density calculated only from the main valence band: a) charge density of $SrPb_3$ calculated in the [111] plane, in which kagome-like lattice is visible; b) charge density of $Sr_{0.5}Ba_{0.5}Pb_3$ calculated in XY planes at various *z* to show kagome-like lattice and triangle lattice. c) charge density of $SrPb_3$ calculated like in (b) and in the YZ plane to show that layers of atoms are not separated from each other.

Fig. 1

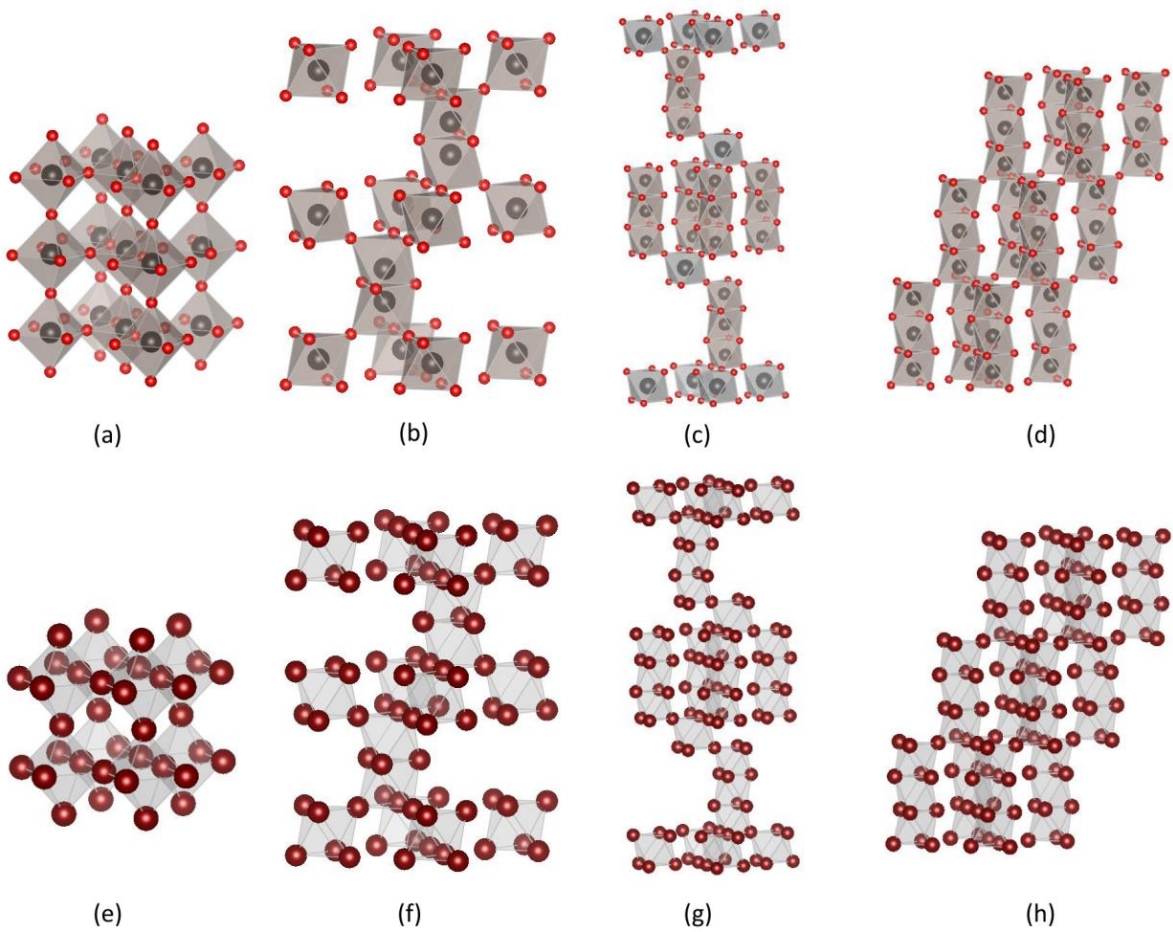

Fig. 2

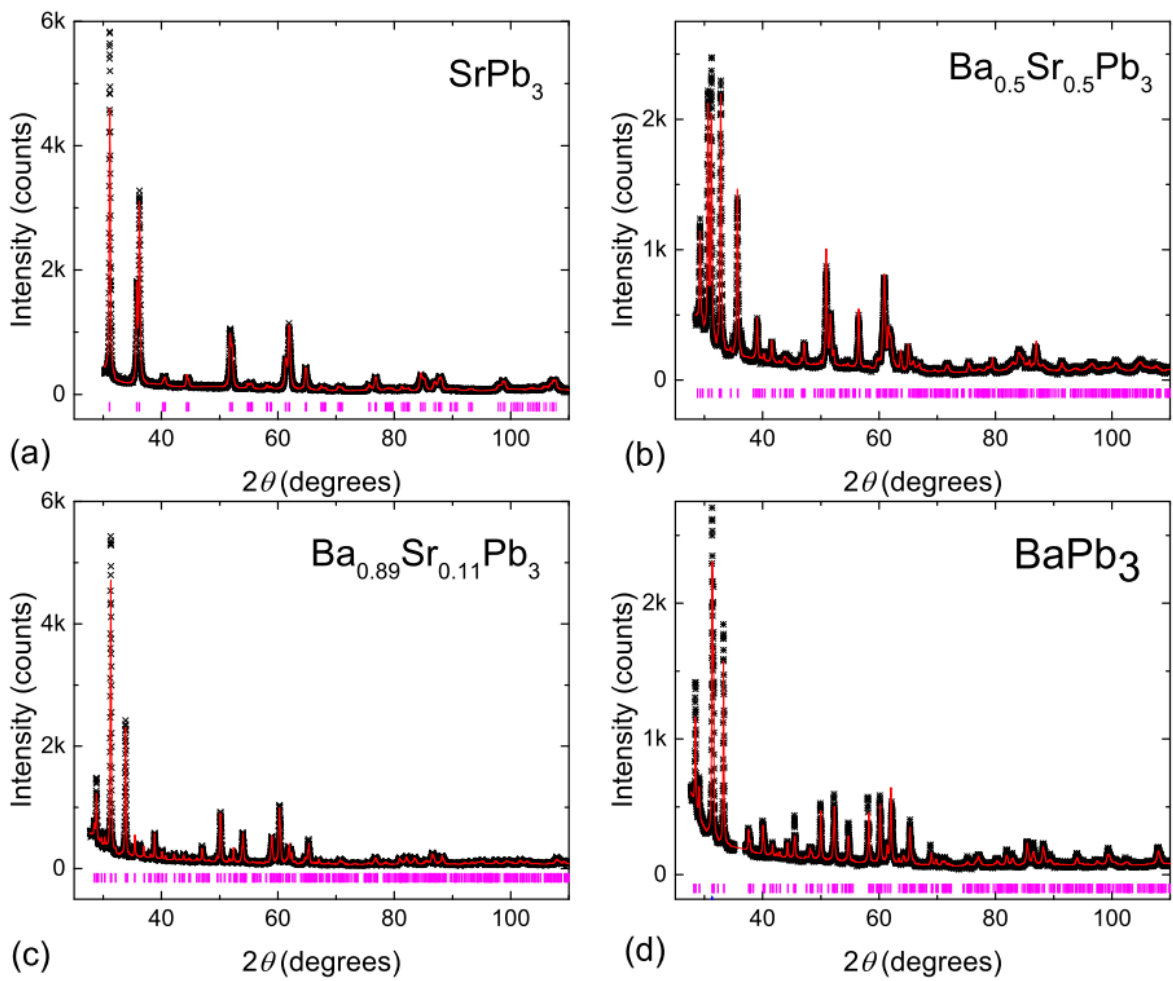

Fig. 3

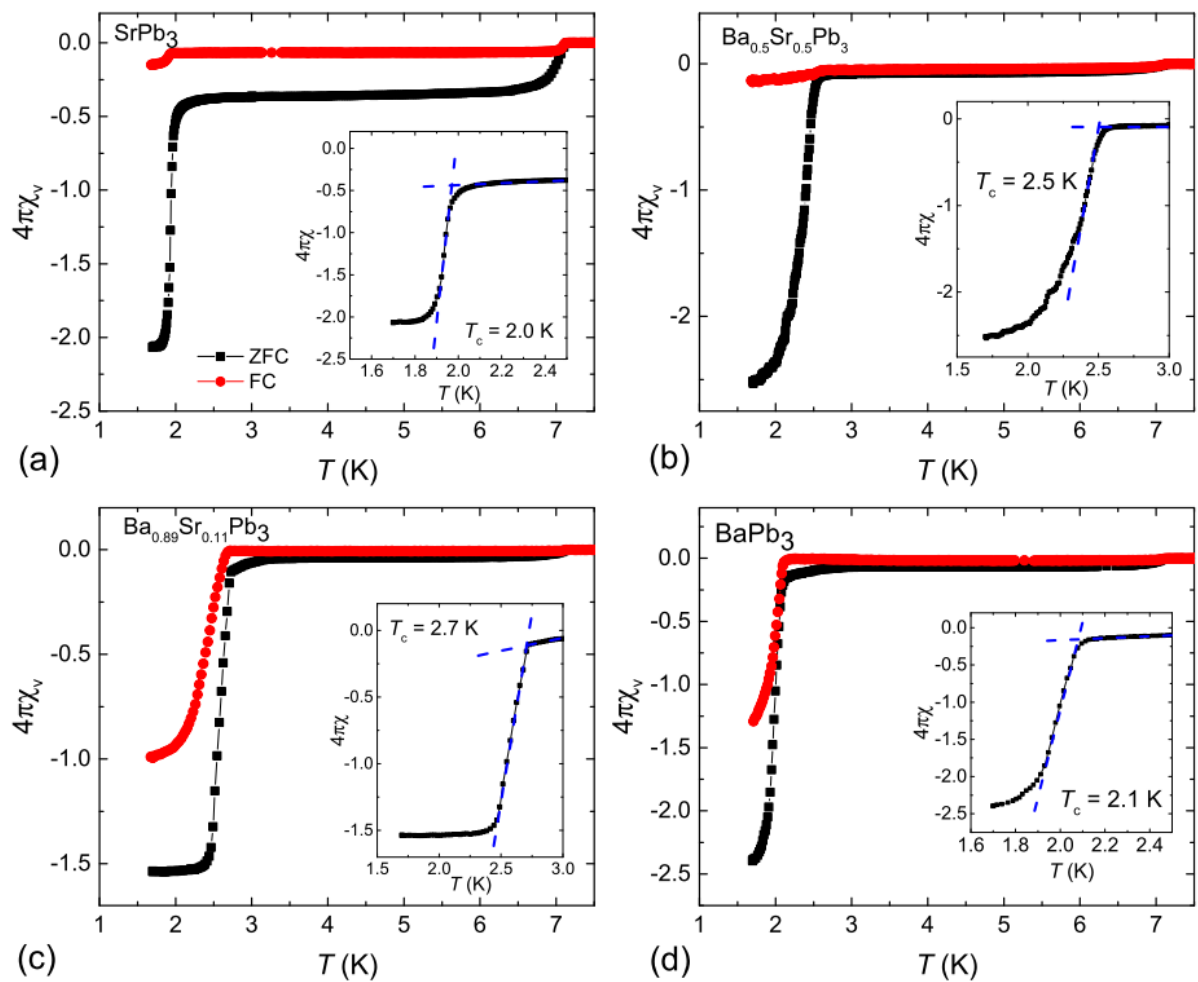

Fig. 4

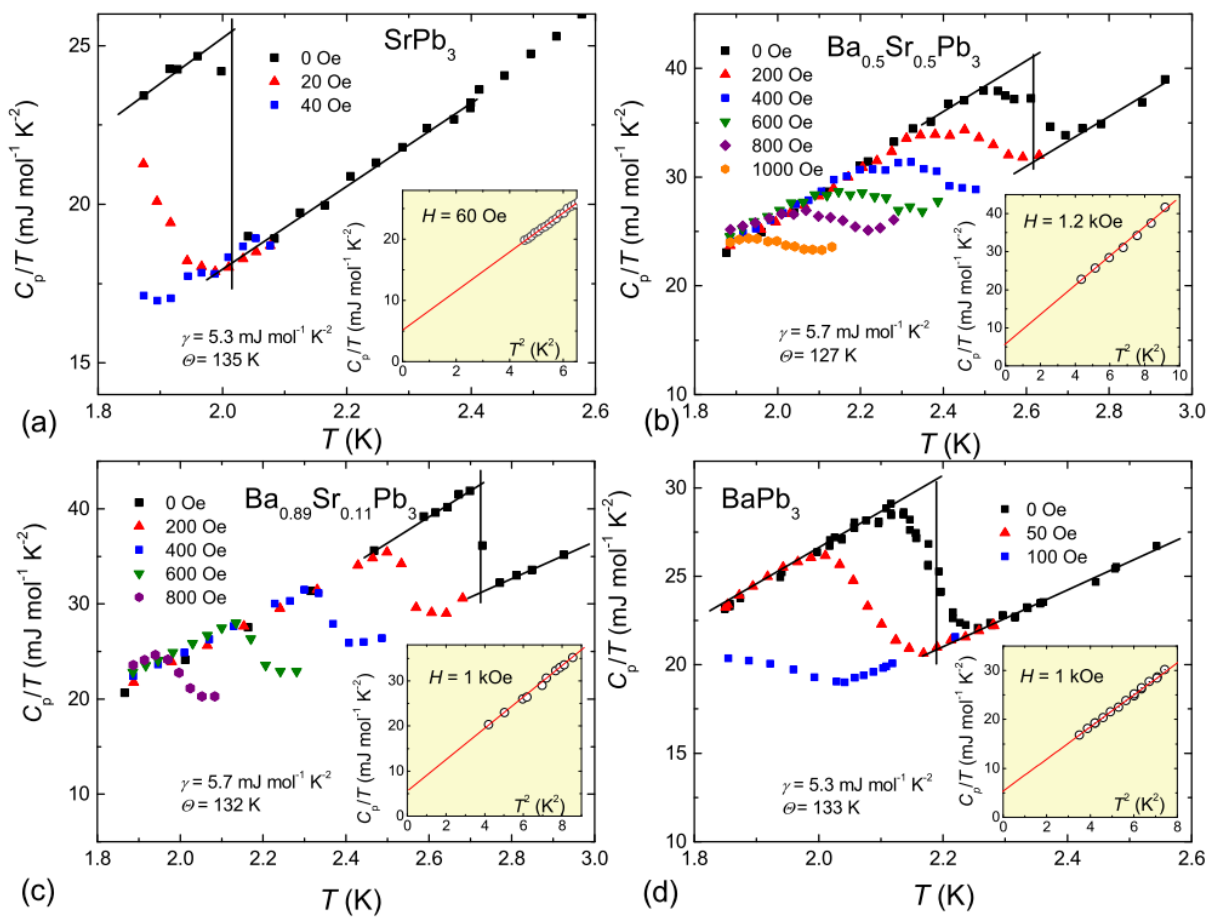

Fig. 5

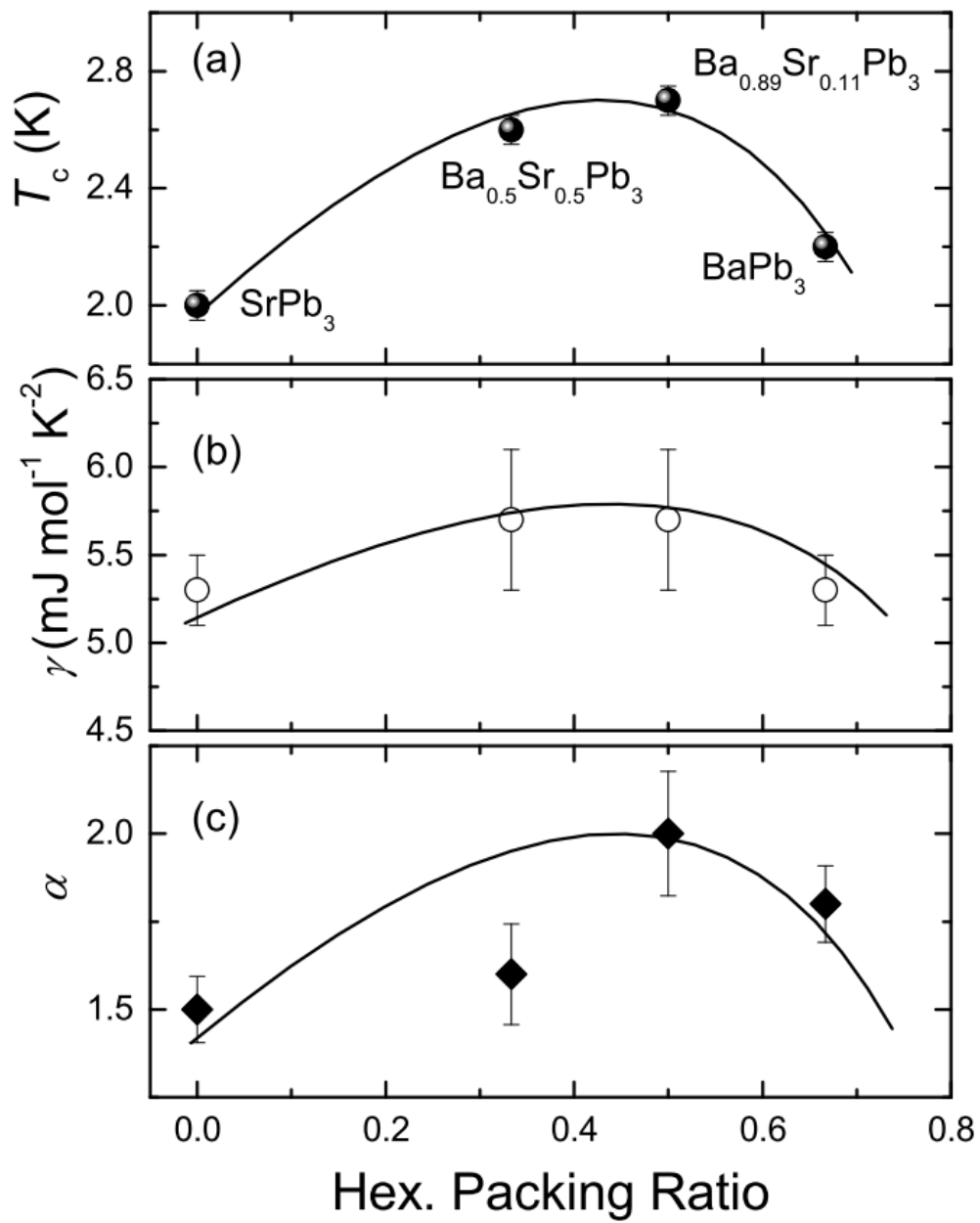

Fig. 6

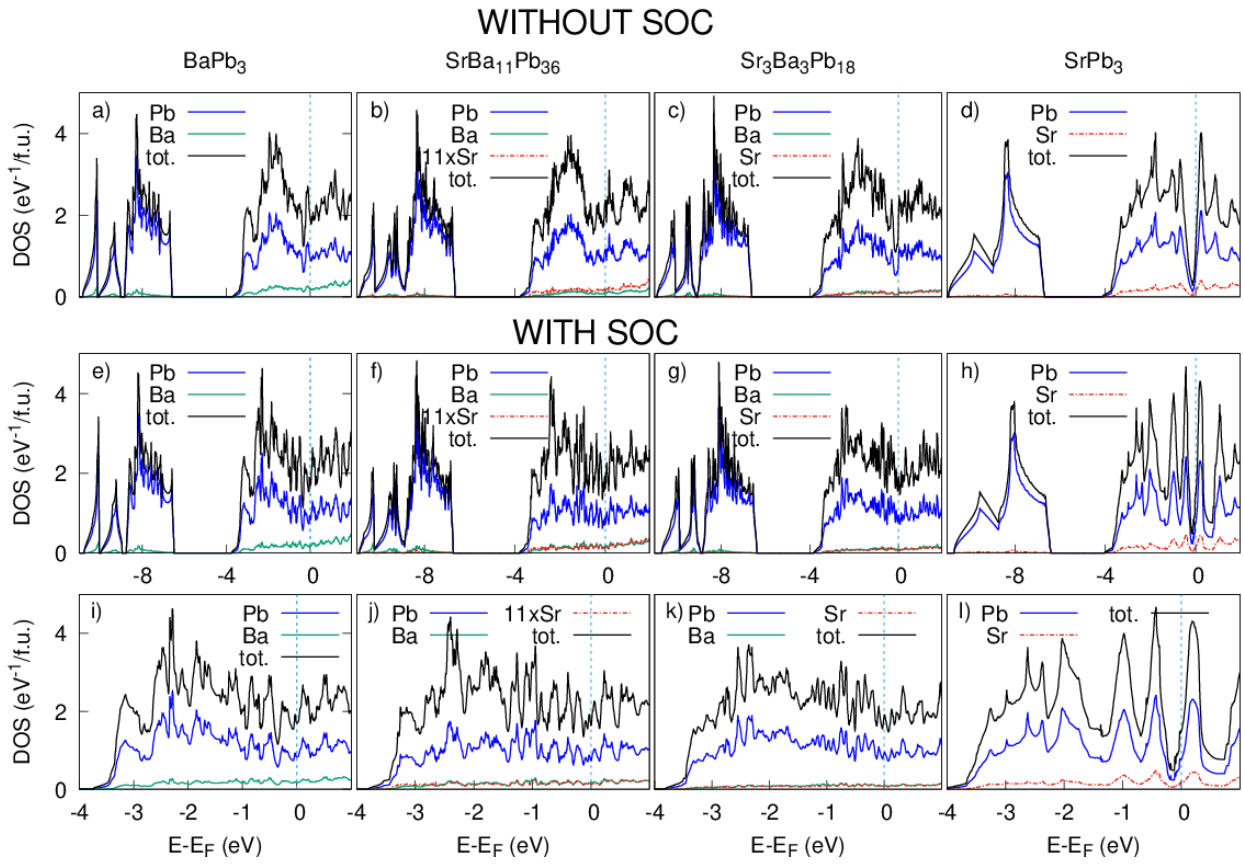

Fig. 7

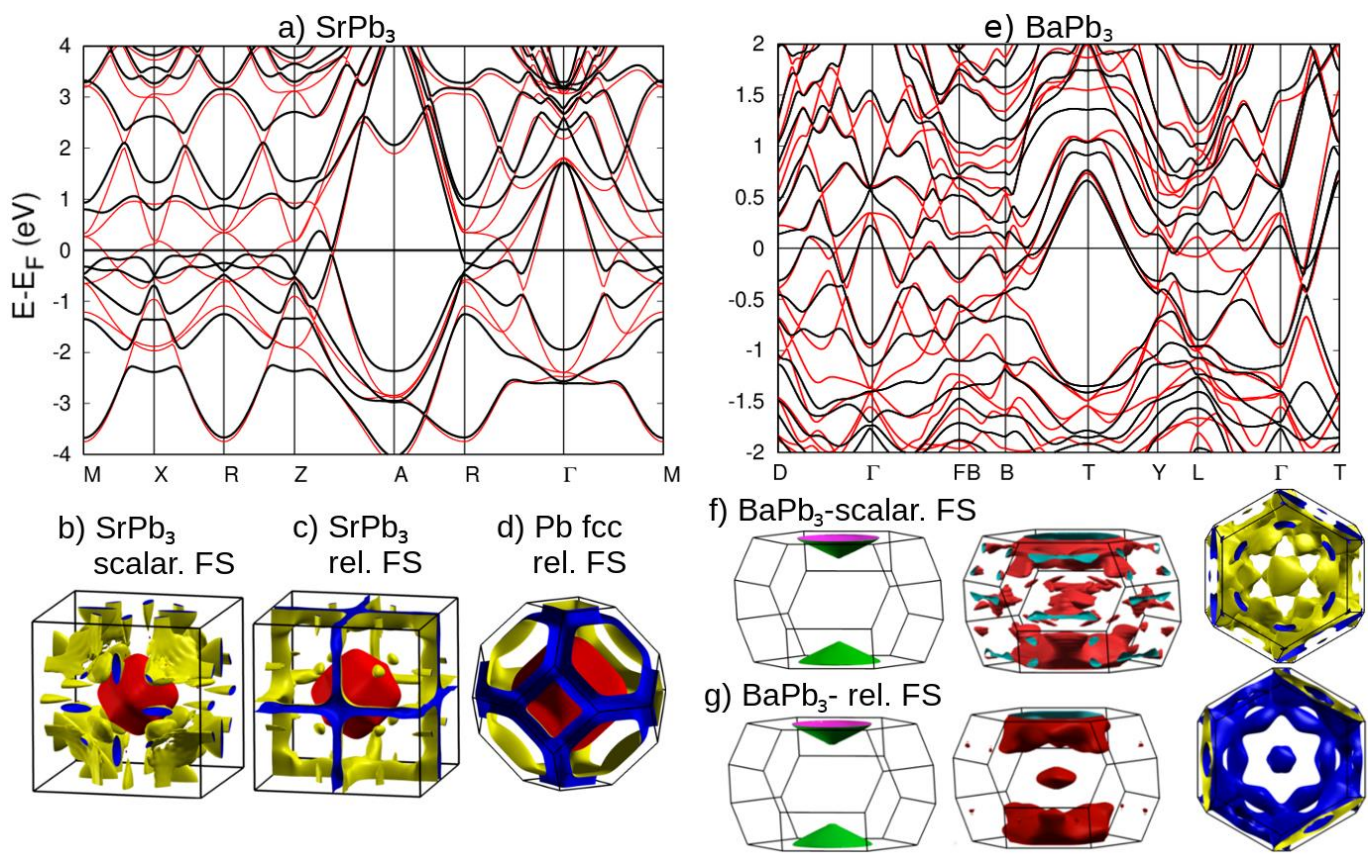

Fig. 8

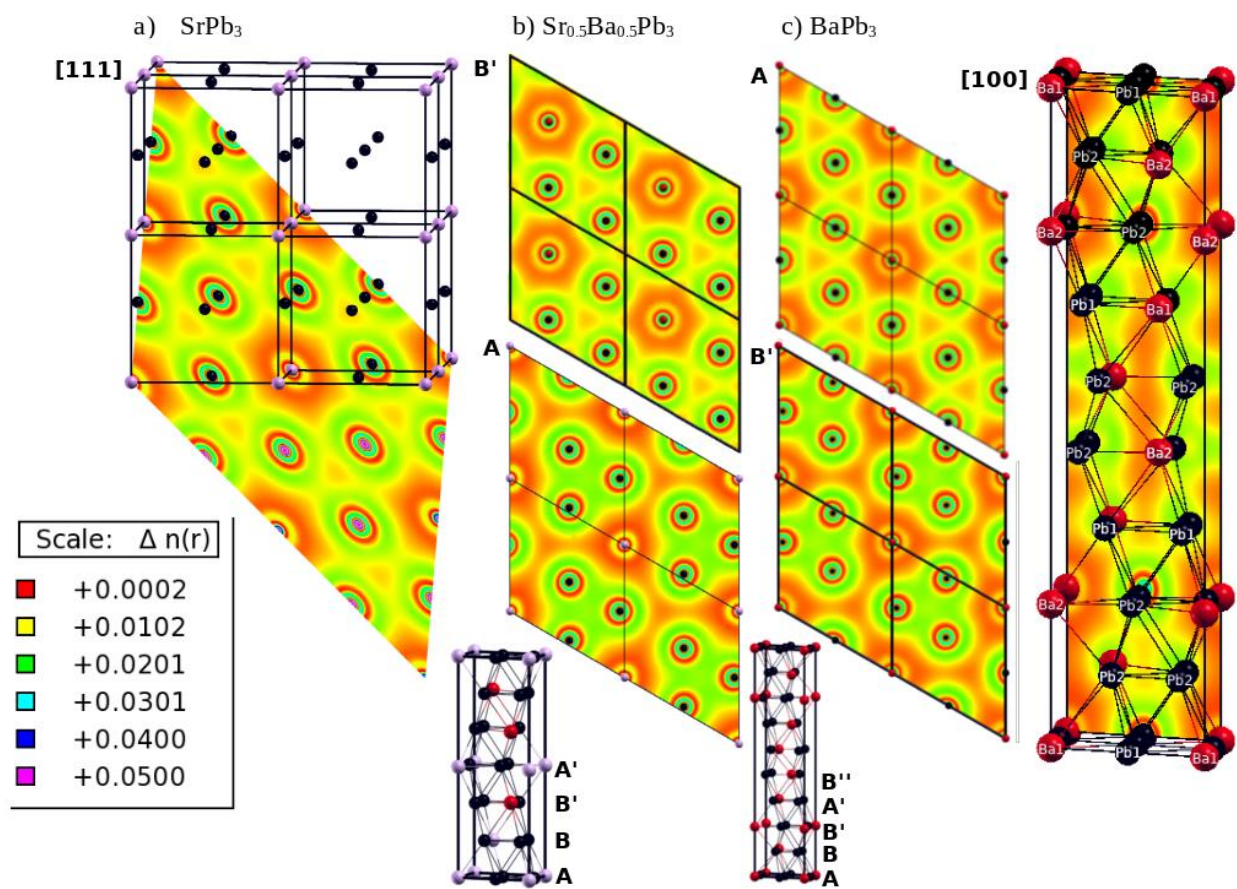